\documentclass{article}
\usepackage{spconf,amsmath,graphicx}

\usepackage{hyperref}       
\usepackage{url}            
\usepackage{booktabs}       
\usepackage{amsfonts}       
\usepackage{xcolor}         
\usepackage{amssymb}
\usepackage{makecell}

\newcommand{\ie}{\textit{i}.\textit{e}.}
\newcommand{\eg}{\textit{e}.\textit{g}.}

\usepackage{pifont}

\usepackage{arydshln}
\usepackage{wrapfig}
\usepackage{multirow}

\makeatletter
\def\blfootnote{\xdef\@thefnmark{}\@footnotetext}
\makeatother

\title{Lip-to-Speech Synthesis in the Wild with Multi-task Learning}
\name{Minsu Kim$^*$, Joanna Hong$^*$\thanks{This work was supported by the National Research Foundation of Korea (NRF) grant funded by the Korea government (MSIT) (No. NRF-2022R1A2C2005529). $^*$Equal Contribution. $^\dagger$Corresponding Author.}, Yong Man Ro$^\dagger$}
\address{School of Electrical Engineering, KAIST, South Korea}
\begin{document}
%
\maketitle
\begin{abstract}
Recent studies have shown impressive performance in Lip-to-speech synthesis that aims to reconstruct speech from visual information alone. However, they have been suffering from synthesizing accurate speech in the wild, due to insufficient supervision for guiding the model to infer the correct content. Distinct from the previous methods, in this paper, we develop a powerful Lip2Speech method that can reconstruct speech with correct contents from the input lip movements, even in a wild environment. To this end, we design multi-task learning that guides the model using multimodal supervision, \textit{i.e.} text and audio, to complement the insufficient word representations of acoustic feature reconstruction loss. Thus, the proposed framework brings the advantage of synthesizing speech containing the right content of multiple speakers with unconstrained sentences. We verify the effectiveness of the proposed method using LRS2, LRS3, and LRW datasets. 

\end{abstract}
\begin{keywords}
Lip-to-speech synthesis, Multi-task learning, Multimodal learning, Speech reconstruction, Lip reading
\end{keywords}

\vspace{-0.2cm}
\section{Introduction}
With the recent development of Artificial Intelligence (AI) technology, interest in solving problems by connecting AI and humans is increasing to help human life. It is also necessary in human-to-human conversations in everyday life, especially when the importance of virtual meetings and video conferencing is highlighted. Among many problems in human-to-human conversations, the need for technologies recognizing an accurate conversation when voice signals are hardly available has been increasing. This technology is promising since it can help people understand conversation in situations like crowded shopping mall, party with lots of people and loud music, and silent video conference.

There have been ongoing studies in speech reconstruction and speech recognition with visual information only. However, since they solely depend on the visual information (\ie, lip movements) which holds incomplete information about speech \cite{kim2022cromm}, these techniques are known to be challenging problems. Especially, video-driven speech reconstruction, also known as Lip-to-speech synthesis (Lip2Speech), has been shown much lower performance compared to visual speech recognition \cite{vougioukas2019video,michelsanti2020vocoderbased,kim2021lip,mira2022endgan}. Therefore, Lip2Speech is being developed with constrained datasets \cite{cooke2006grid,harte2015tcd} such that the number of speakers is limited or the sentences following a fixed grammar. In contrast, visual speech recognition \cite{afouras2018deep,ma2021conformer,kim2022distinguishing} achieved significant performance improvements in the wild datasets \cite{chung2016lrw,chung2017lrs2,afouras2018lrs3} containing large speaker variations and utterances. The performance gap between the two technologies is because Lip2Speech needs to consider the varying characteristics in speech (\textit{e.g.}, voice, accent, and intonation). Also, in Lip2Speech task, the reconstruction criteria of continuous audio representation is insufficient compared to that of visual speech recognition, while visual speech recognition can be trained with discrete supervision of text, unnecessary to consider the speaker characteristics.

To reduce the performance gaps between Lip2Speech and visual speech recognition, in this paper, we develop a powerful Lip2Speech method that can correctly capture spoken words even in wild environments. To this end, we propose a multi-task learning method that learns to predict text with content supervision and learns to predict acoustic features. Therefore, we can complement the insufficient content guidance in acoustic feature reconstruction criteria through the proposed content supervision. Specifically, we propose two different types of content supervision: feature- and output-level. In the feature-level content supervision, the model is guided to predict aligned text from the input visual representations with Connectionist Temporal Classification (CTC) loss \cite{graves2006ctc} before synthesizing the acoustic features. Therefore, we can bring strong supervision of text into Lip2Speech without losing the alignment of visual inputs and audio outputs. In the output-level content supervision, we adopt an Automatic Speech Recognition (ASR) model to feedback the model to synthesize speech containing the correct words. Along with the proposed content supervision, the model is guided by reconstruction loss which is applied at continuous auditory features, thus forming multi-task learning. Hence, through the proposed framework, it is possible to synthesize high-quality speech by watching only the visual information even in the wild datasets which are composed of utterances from hundreds of speakers and unconstrained sentences. 

To validate the effectiveness of the proposed method, we utilize LRS2 \cite{chung2017lrs2} and LRS3 \cite{afouras2018lrs3}, the largest sentence-level audio-visual datasets obtained in the wild, and LRW \cite{chung2016lrw}, a word-level dataset. Through comprehensive experiments, we show that the synthesized speech of the proposed method contains correct contents with the lowest Word Error Rates (WERs) compared to the previous state-of-the-art methods.

\vspace{-0.2cm}
\section{RELATION TO PRIOR WORK}
Early works \cite{akbari2018lip2audspec,vougioukas2019video} utilized constrained datasets to develop the Lip2Speech models. Later, \cite{l2w,mira2022endgan,kim2021lip,hong2021speech} improved the architecture and training methods, and provided the possibility of unconstrained Lip2Speech. Recently, \cite{mira2022svts} utilized the LRS3 dataset which has no restriction in both the number of speakers and sentences. However, they failed to successfully measure WER for the LRS3 dataset due to difficulties in synthesizing accurate speech in unconstrained sentence-level dataset. Different from the previous works, in this paper, we focus on synthesizing speech with accurate contents by proposing content supervisions at different levels.

\vspace{-0.2cm}
\section{Method}
\begin{figure}[t]
	\begin{minipage}[b]{1.0\linewidth}
		\centering
		\centerline{\includegraphics[width=7.0cm]{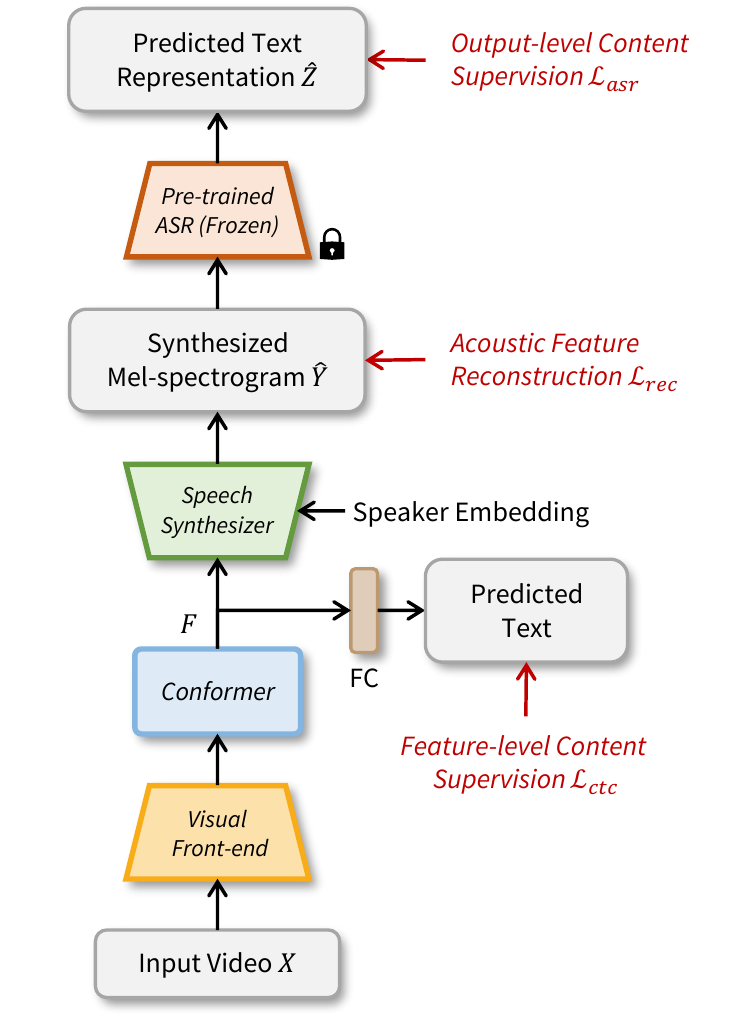}}
	\end{minipage}
	\vspace{-0.4cm}
	\caption{Overview of the proposed multi-task learning.}
	\label{fig:1}
	\vspace{-0.4cm}
\end{figure}

Let $X=\{x_1,\dots,x_T\}\in\mathbb{R}^{T\times H\times W\times C}$ be an input video containing lip movements, $Y=\{y_1,\dots,y_S\}\in\mathbb{R}^{K\times S}$ be acoustic feature of ground-truth speech constructed with mel-spectrogram, and $U=\{u_1,\dots,u_L\}\in\mathbb{R}^{L}$ be the ground-truth transcription of the utterance with $L$ tokens. Here, $T$ indicates the frame lengths, $H$, $W$, and $C$ are the frame height, width, and channel sizes, respectively, $K$ and $S$ are the mel-spectral channel and the sequence length, respectively, and $L$ represents the length of the transcription. Our main goal is to translate the input lip video $X$ into adequate speech $Y$ without constraint on the number of speakers or sentences (\ie, in the wild). To this end, we guide the model with multi-tasks, text prediction using content supervision and acoustic feature prediction using reconstruction supervision. For the content supervision, we employ two different types of supervision, feature- and output-level. The details of the aforementioned criteria will be described in the following subsections. The overall proposed framework is illustrated in Fig. \ref{fig:1}.

\subsection{Feature-level content supervision}
To synthesize speech by watching the lip movements only, predicting the right spoken words in advance is important. However, most previous works \cite{akbari2018lip2audspec,vougioukas2019video,mira2022endgan,l2w,kim2021lip} depend on the reconstruction loss (\eg, L1 and L2 loss) between the predicted and ground-truth acoustic representations (\eg, MFCC or mel-spectrogram). Different from the discrete nature of text, meaning that the same words always have the same labels, acoustic representations might contain different values, based upon the different characteristics of the speakers, tones, accents, and so on, for the same words \cite{van2017neuraldiscrete}.
Thus, using only reconstruction loss for continuous acoustic features for Lip2Speech can be insufficient for guiding the network to infer the right words from the silent talking face video. In order to mitigate this problem, we additionally utilize the discrete text modality with Connectionist Temporal Classification (CTC) loss \cite{graves2006ctc}. Since CTC loss not only guides the network to infer the right words but also produces the aligned representations, it is adequate for the Lip2Speech task which should maintain the visual-audio synchronization.

The frames of input video $X$ is embedded with CNN-based visual front-end $\Phi_v$ and their temporal relationships are modeled by conformer $\Phi_c$ \cite{gulati2020conformer}, as follows,
\begin{align}
    F = \Phi_c(\Phi_v(X)).
\end{align}
The final encoded visual representations $F=\{f_1,\dots,f_T\}\in\mathbb{R}^{T\times D}$ are employed for both text prediction and speech prediction, where $D$ is the embedding size. The text prediction $p = \text{Softmax}(FW_{ctc} + b_{ctc})$ is guided with the CTC loss defined as $\mathcal{L}_{ctc}(p,U)$, where $W_{ctc}\in\mathbb{R}^{D\times N}$ and $b_{ctc}\in\mathbb{R}^N$ are the weight and bias for text prediction, respectively, and $N$ refers to the number of classes. With the feature-level content supervision, the visual representations $F$ can contain the correct wordings of the input visual speech, which will be eventually translated into output acoustic features.

\subsection{Output-level content supervision}
As we discussed in the previous section, using CTC loss before synthesizing the speech can explicitly guide the model to learn the correct words. In addition, along with the feature-level content supervision, we can also impose content supervision at the output level. To this end, we propose to bring a feedback network that guides the model to focus on modelling accurate content even when the text annotation is unavailable. We utilize a pre-trained ASR model to examine whether the synthesized speech contains correct words or not. The content representations $\hat{Z}$ of synthesized speech $\hat{Y}$ are extracted from the ASR model before the last classification layer. Then, the content representations $\hat{Z}$ are guided to resemble with the ground-truth content representations $Z$ which is extracted from the ground-truth speech $Y$ as follows,
\begin{gather}
    Z = \text{ASR}(Y), \quad \hat{Z} = \text{ASR}(\hat{Y}), \\
    \mathcal{L}_{asr} = ||Z - \hat{Z}||_2^2.
\end{gather}
Therefore, the model can focus more on modelling accurate content in the output with the help of the feedback model. Moreover, since the output-level content supervision does not require any text annotation, we can employ it even if the text annotations are unavailable.

\vspace{-0.3cm}
\subsection{Lip-to-Speech synthesis in the wild}
Different from the previous work, \cite{mira2022svts}, we put the speaker embedding \cite{l2w}, which for providing the speaker characteristics, right before the speech synthesizer so that the front models (\ie, visual front-end and conformer) can focus more on modelling speech content. Moreover, to properly embed the speaker characteristics into the final output speech, we use a deeper speech synthesizer, composed of 1D CNN layers.

From the visual features $F$, acoustic features are generated by Speech Synthesizer $\Psi$. The generation is guided with the reconstruction loss as follows,
\begin{align}
    \mathcal{L}_{rec} = || \Psi(F) - Y ||_1.
\end{align}
By weighted summing the three loss functions defined, the objective function for the proposed multi-task learning can be formulated as follows, $\mathcal{L}_{tot}= \lambda_{ctc} \mathcal{L}_{ctc} + \lambda_{asr} \mathcal{L}_{asr} + \lambda_{rec} \mathcal{L}_{rec}$, where $\lambda$ are the balancing weights. By guiding the network with multimodal supervision, the synthesized speech $\hat{Y}=\Psi(F)$ can contain accurate contents with the help of content supervision. For stable training, we use 0 for $\lambda_{asr}$ for early epochs and turn on the output-level content supervision when the reconstruction loss falls enough.

\vspace{-0.1cm}
\section{Experiments}
\vspace{-0.1cm}
\subsection{Datasets}
\vspace{-0.1cm}
\noindent {\bf LRS2} \cite{chung2017lrs2} dataset is utilized to validate the performance of the proposed method in the wild environment without constraints of the number of speakers and sentences. It has about 142,000 utterances including pre-train and train sets. We utilize both sets for training (about 223 hours), and test the model on a test set containing 1,243 utterances.

\noindent {\bf LRS3} \cite{afouras2018lrs3} dataset is another large-scale audio-visual corpus dataset. We follow the unseen data splits of \cite{mira2022svts}. For training, about 131,000 utterances are utilized (about 296 hours), and 1,308 utterances are used for testing.

\noindent {\bf LRW} \cite{chung2016lrw} dataset is a word-level English audio-visual corpus. It contains 500 word classes with a maximum of 1,000 training videos each. The dataset is collected from BBC news program, thus containing large speaker, pose, and illumination variations. Compared to the previously mentioned sentence-level datasets in the wild, it is constrained to 500 words. The total training data length is about 157 hours.

\vspace{-0.1cm}

\subsection{Implementation details}
\subsubsection{Dataset preprocessing}
\vspace{-0.2cm}
For all datasets, we crop the lip regions, resize the cropped frames into 112$\times$112, and convert the colors into grayscale as \cite{kim2021lip}. Audio with a sample rate of 16kHz is converted to mel-spectrogram by using a hop length of 10ms and window length of 40ms. For the data augmentation, we randomly erase the spatial region of the video consistently in all frames, and time-masking \cite{mira2022svts} is applied to the input video which is beneficial in modelling visual context. For the text tokenizer, we employ sentencepiece \cite{kudo2018sentencepiece} for the LRS2 and LRS3 with 4,000 subwords, and word class for the LRW dataset.

\vspace{-0.2cm}
\subsubsection{Architectural details}
\vspace{-0.2cm}
For the visual front-end, ResNet18 whose first layer is modified with a 3D convolution is utilized \cite{petridis2018end}. For the conformer, we use 12 layers, 8 heads, and convolution kernel size of 31, for the LRS2 and LRS3. For the LRW, we use 6 layered conformer. For the speech synthesizer, we use 3 layered 1D convolutions with 256, 128, and 320 hidden dimensions and a kernel size of 7, and the speaker embedding \cite{l2w,mira2022svts} is obtained by encoding random short audio clip of the same speaker with another 3 layered 1D convolutions with 128, 256, and 256 hidden dimensions and a kernel size of 7. For the LRW, we use random 0.2 sec of audio clip and 0.5 sec for both LRS2 and LRS3. The speaker embedding setting is different from the previous work \cite{mira2022svts} that utilized all audio frames to extract the speaker embedding by using a pre-trained speaker verification model on large-scale extra datasets. For each LRS2 and LRS3 dataset, the pre-trained ASR model is trained using CTC loss and is composed of 6 conformer encoders with 4 attention heads. For the LRW dataset, we utilize temporal average pooling and cross entropy loss. The pre-trained ASR models keep being frozen during training the Lip2Speech model.

\vspace{-0.2cm}
\subsubsection{Training details}
\vspace{-0.2cm}
For training, we use an initial learning rate of 0.0001, batch size of 16, and AdamW \cite{loshchilov2017decoupled} optimizer. For the LRS2 and LRS3 datasets, mel-spectrogram of 200 frames are generated by randomly selecting corresponding visual representations from $F$, to save computing memory. $\lambda_{ctc}$, $\lambda_{asr}$, and $\lambda_{rec}$ are empirically set to 1, 1, and 100, respectively.

\begin{table}[t]
	\renewcommand{\arraystretch}{1.2}
	\renewcommand{\tabcolsep}{2.0mm}
\caption{Ablation study on LRS2 dataset. `C.S.' is an abbreviation for content supervision}
\centering
\resizebox{0.99\linewidth}{!}{
\begin{tabular}{lcccc}
\hline
\textbf{Method} & \textbf{STOI} & \textbf{ESTOI} & \textbf{PESQ} & \textbf{WER}(\%) \\ \hline
Baseline \cite{mira2022svts} & 0.491 & 0.291 & 1.34 & 93.91 \\ 
\,\, $+$ Feature-level C.S. & 0.513 & 0.312 & 1.34 & 76.40 \\ 
\,\, $+$ Conv1D & 0.518 & 0.320 & 1.34 & 75.32 \\ 
\,\, $+$ Speaker Embedding & 0.519 & 0.329 & \textbf{1.36} & 68.80 \\
\,\, $+$ Output-level C.S. (\textbf{Full}) & \textbf{0.526} & \textbf{0.341} & \textbf{1.36} & \textbf{60.54} \\ \hdashline
\,\, $-$ Feature-level C.S. & 0.496 & 0.299 & 1.34 & 84.27 \\
\hline
\end{tabular}}
\label{table:1}
\vspace{-0.2cm}
\end{table}

\vspace{-0.2cm}
\subsubsection{Evaluation metrics}
\vspace{-0.2cm}
For evalutation, we use Short-Time Objective Intelligibility (STOI) \cite{stoi} and Extended STOI (ESTOI) \cite{estoi} to measure the intelligibility of generated speech, Perceptual Evaluation of Speech Quality (PESQ) \cite{pesq} to measure the perceptual quality of generated speech, and Word Error Rate (WER) to measure how the generated speech containing the correct contents. Note that we focus on WER metric to examine whether the proposed method is effective in modelling speech content. The quantitative results are obtained by converting into the waveform with Griffin-Lim algorithm \cite{griffinlim}. Note that the ASR model used to train the Lip2Speech model and that used to measure quantitative results are different.

\begin{table}[t]
	\renewcommand{\arraystretch}{1.2}
	\renewcommand{\tabcolsep}{3.5mm}
\caption{Performance comparisons on LRS2 dataset}
\centering
\resizebox{0.99\linewidth}{!}{
\begin{tabular}{ccccc}
\hline
\textbf{Method} & \textbf{STOI} & \textbf{ESTOI} & \textbf{PESQ} & \textbf{WER}(\%) \\ \hline
VCA-GAN \cite{kim2021lip} & 0.407 & 0.134 & 1.24 & 109.01 \\
SVTS \cite{mira2022svts} & 0.491 & 0.291 & 1.34 & 93.91 \\ 
\textbf{Proposed Method} & \textbf{0.526} & \textbf{0.341} & \textbf{1.36} & \textbf{60.54} \\ 
\hline
\end{tabular}}
\label{table:4}
\vspace{-0.3cm}
\end{table}

\subsection{Ablation study}
We examine the effectiveness of the proposed method by adding each component to the baseline, SVTS \cite{mira2022svts}. We use the LRS2 dataset to validate the efficacy of the proposed framework in the wild environment. Table \ref{table:1} shows the ablation results. The WER of baseline model is 93.91\%, meaning that the baseline model can hardly generate speech with correct content. By using the feature-level content supervision (+ Feature-level C.S), we improve the WER by 17.51\% which is a large improvement compared to the baseline model. Moreover, by changing the speech synthesizer with a deeper architecture (+ Conv1D) and the injection location of speaker embedding (+ Speaker Embedding), the performance is improved to 68.80\% WER. Finally, with the proposed output-level content supervision (+ Output-level C.S) which is the final proposed method (\ie, Full), we achieve 60.54\% WER, and the proposed model outperforms the baseline model by 33.37\% WER. Finally, in the case of that text annotation is not available (- Feature-level C.S), we achieve 84.27\% WER with the output-level content supervision, which improves the WER by about 10\% from the baseline.

The final performance (60.5\% WER) is significant since a well-known visual speech recognition with CTC architecture \cite{afouras2018deep} achieves about 65\% WER on LRS2 dataset. This means that our proposed Lip2Speech model achieves comparable performance with the visual speech recognition tasks.

\begin{table}[t]
	\renewcommand{\arraystretch}{1.2}
	\renewcommand{\tabcolsep}{3.5mm}
\caption{Performance comparisons on LRS3 dataset}
\centering
\resizebox{0.99\linewidth}{!}{
\begin{tabular}{ccccc}
\hline
\textbf{Method} & \textbf{STOI} & \textbf{ESTOI} & \textbf{PESQ} & \textbf{WER}(\%) \\ \hline
VCA-GAN \cite{kim2021lip} & 0.474 & 0.207 & 1.23 & 96.63 \\
SVTS \cite{mira2022svts} & \textbf{0.507} & \textbf{0.271} & 1.25 & 79.83\\ 
\textbf{Proposed Method} & 0.497 & 0.268 & \textbf{1.31} & \textbf{66.78} \\ 
\hline
\end{tabular}}
\label{table:5}
\vspace{-0.3cm}
\end{table}

\blfootnote{$^\ddagger$Reported using Google API.}

\subsection{Comparisons with previous methods}
We compare the performances with the previous methods by using the large-scale audio-visual datasets, LRS2 and LRS3, composed of various utterances from diverse speakers.
Since there is no prior work that tried to measure the WER on such wild sentence-level datasets, we train VCA-GAN \cite{kim2021lip} and SVTS \cite{mira2022svts} on the LRS2 and measure the WER, shown in Table \ref{table:4}.
Compared to the previous methods, VCA-GAN and SVTS, the results confirm that the proposed method synthesizes the speech with accurate wordings by achieving the best WER and PESQ. Moreover, in Table \ref{table:5}, we obtain consistent results for the LRS3 dataset compared to the previous experiments with the large gap in the WER performances. 

In addition, we verify that our proposed model achieves comparable performances with the other popular methods using word-level dataset, LRW. As indicated in Table \ref{table:3}, the proposed method achieves comparable performances with the state-of-the-art methods showing that multi-task learning can be also applied to the word-level dataset.

\begin{table}[t]
	\renewcommand{\arraystretch}{1.2}
	\renewcommand{\tabcolsep}{3mm}
\caption{Performance comparisons on LRW dataset.}
\centering
\resizebox{0.99\linewidth}{!}{
\begin{tabular}{ccccc}
\hline
\textbf{Method} & \textbf{STOI} & \textbf{ESTOI} & \textbf{PESQ} & \textbf{WER}(\%) \\ \hline
Lip2Wav \cite{l2w} & 0.543 & 0.344 & 1.20 & 34.20$^\ddagger$ \\
VCA-GAN \cite{kim2021lip} & 0.565 & 0.364 & 1.34 & 32.07\\
End-to-end GAN \cite{mira2022endgan} & 0.552 & 0.330 & 1.33 & 41.31\\
SVTS \cite{mira2022svts} & \textbf{0.649} & \textbf{0.483} & 1.49 & \textbf{12.53} \\
\textbf{Proposed Method} & 0.642 & 0.476 & \textbf{1.56} & 13.86 \\ 
\hline
\end{tabular}}
\label{table:3}
\vspace{-0.2cm}
\end{table}

\vspace{-0.2cm}
\section{CONCLUSION}
In this paper, we design a powerful Lip2Speech framework that works in the wild. To this end, we propose multi-task content learning: feature- and output-level content supervisions, with the acoustic feature reconstruction. The extensive experimental results prove that the proposed learning framework effectively translates the lip movements into speech audio with the accurate content, even in the wild environments.

{\small
\bibliographystyle{IEEEbib}
\bibliography{egbib}
}

\end{document}